\documentclass[12pt,preprint]{aastex}

\usepackage{graphicx}

\def\lsim{\hbox{ \rlap{\raise 0.425ex\hbox{$<$}}\lower 0.65ex\hbox{$\sim$} }}
\def\gsim{\hbox{ \rlap{\raise 0.425ex\hbox{$>$}}\lower 0.65ex\hbox{$\sim$} }}

\def\opeqn{\begin{equation}}
\def\cleqn{\end{equation}}

\def\dint{\displaystyle{\int}}
\def\dfrac{\displaystyle\frac}

\pdfoutput = 1

\begin{document}

\title{Elliptically Symmetric Lenses and Violation of Burke's Theorem}
\author{Sun Hong Rhie }
\affil{Physics Department, University of Notre Dame, IN 46556}
\email{eplusminus@gmail.com}


\begin{abstract}
We show that the outside equation
of a bounded elliptically symmetric lens (ESL)  
exhibits a pseudo-caustic that arises from a branch cut.
A pseudo-caustic is a curve in the source 
plane across which the number of images changes by one.
The inside lens equation of a bounded ESL is free of a 
pseudo-caustic. 
Thus the total parity of the images of a point source lensed
by a bounded elliptically symmetric mass is not an invariant 
in violation of the Burke's theorem. A smooth mass density
function does not guarantee the validity of the Burke's theorem.

Pseudo-caustics of various lens equations are discussed.
In the Appendix, Bourassa and Kantowski's deflection angle formula for
an elliptically symmetric lens is reproduced using the Schwarz function 
of the ellipse for an easy access; the outside and inside lens
equations of an arbitrary set of truncated circularly 
or elliptically symmetric lenses, represented as points, sticks, and 
disks, are presented as a reasonable
approximation of the realistic galaxy or cluster lenses. 

One may consider
smooth density functions that are not bounded but fall
sufficiently fast asymptotically to preserve the total parity invariance.
Any bounded function may be sufficiently closely approximated by 
an unbounded smooth function obtained by truncating its Fourier integral
at a high frequency mode. Whether to use a bounded
function or an unbounded smooth function 
for an ESL lens mass density, 
whereby whether to observe the total parity invariance
or not, incurs philosophical questions. 
For example, is it sensible to insist that the elliptical symmetry
of an elliptic lens galaxy be valid in the entire sky?
How a pseudo-caustic
close to or intersecting with a caustic must be withered away
during a smoothing process and what it means will be 
investigated in a separate work.
\end{abstract}


\clearpage

\section{Introduction}

Burke's theorem \citep{burke81}
is known as odd number theorem 
\citep{mackenzie,SEF}, 
and is made
of two pieces of claims. For a bounded mass distribution lens, 1) the total
parity of the images is an invariant and 2) the number of images of a
source sufficiently far away from the lens system is one. 
Here we will only focus on the invariance of the total parity.

Thirty five years ago, \citet{bourassa} found  that the deflection 
angle of an elliptically symmetric lens (ESL) can be expressed as an 
integral over the interval $[0,1]$. In our notations described in the 
Appendix, the normalized lens equation of an ESL is written as
\opeqn
 \omega = z 
  - \dint_0^{t_{cut}} \dfrac{\tilde\sigma(t) t dt}{(\bar z^2-c^2t^2)^{1/2}} 
\label{eqLeqMain}
\cleqn
where 
$\omega$ and $z$ are the positions of the source and an 
image, $c$ is the focal length of the mass distribution ellipse, 
and $t$ is the elliptic radial parameter such that the position
variable of the mass distribution is given by 
$\xi = t(a\sin\theta + i b\cos\theta): \ t\in [0,1]$ where $a$ and $b$ 
are the semi-major and semi-minor axes of the mass ellipse.
The mass in the mass ellipse is used for the normalization of
the lens equation.
$\tilde\sigma(t)$ is the projected mass density function re-normalized
such that 
\opeqn
 1 =  \int_0^1 \tilde\sigma(t) t dt \ .
\cleqn
If the image position $z$ is inside the mass ellipse, then 
$t_{cut} = \eta$ where $\eta$ is the elliptic radial parameter of $z$:
$z = \eta(a\sin\theta + i b\cos\theta)$,
and if the image position $z$ is outside the mass ellipse, then
$t_{cut} = 1$.

A pseudo-caustic refers to a curve
in the source plane across which the number of images changes 
by one and was considered a curious phenomenon \citep{evans98}.
The appearance of a pseudo-caustic is an indicator that the
total parity of the images is not an invariant in violation of the 
Burke's theorem.
The integrand of eq.(\ref{eqLeqMain}) includes a square root
function,  and we will see that that is the source of the pseudo-caustic 
of an ESL outside lens equation. 

We adopted the terminology pseudo-caustic because 1) a 
pseudo-caustic curve can form cusps in conjunction with open 
caustic curves and 2) we can continue to use the term ``caustic
domains" to describe the domains of the source plane with 
definite number of images that are defined not only by caustics
but also by pseudo-caustics.

In section \ref{secTwo} we discuss the pseudo-caustics of various
lens equations. The outside equation of a generic
elliptically symmetric lens is shown to have a pseudo-caustic 
that arises from a branch cut. The inside equation of an elliptically
symmetric finite density lens does not have a pseudo-caustic.
Thus a bounded elliptically symmetric lens violates the invariance
of the total parity of the images; the arguments of smooth density
functions for Burke's theorem are invalidated. 
In section \ref{secThree} we give a summary and
raise a possibility to approximate a bounded mass density
by a smooth function with infinite extension
to preserve the invariance of the total
parity of the images. We leave the details to a separate work.
It is a natural wonder if other types of singularities than point 
singularities and branch cuts may be to be found
in gravitational lensing.
In the Appendix, we reproduce the deflection angle formula of 
\citet{bourassa} using the Schwarz function of the ellipse 
\citep{FKK07,KL09} and derive various lens equations used in 
section \ref{secTwo}; we also spell out the inside and outside
lens equations of an arbitrary set of circularly symmetric lenses
and elliptically symmetric lenses as a reasonable model for 
galaxy and cluster lenses; they are represented as points,
sticks, and disks.

\section{Pseudo-caustics of Elliptically Symmetric Lens Equations}
\label{secTwo}

We start with examining the singular isothermal lens in which the 
pseudo-caustic arises from a point mass density singularity 
where the Burke's vector field
is ill defined and move on to discussing the pseudo-caustics of
various elliptically 
symmetric lenses (ESL). We will see that the outside lens equation 
of any bounded ESL (mass density vanishes outside a certain
finite radius) has a pseudo-caustic that arises from the
double-valuedness of the c-cut that is from a branch cut.
The inside equation of a bounded ESL does not have a 
pseudo-caustic. Thus a bounded ESL violates the invariance
of the total parity.

\subsection{SIS or Inside a Cylindrically Truncated Isothermal Sphere}

The ``curious behavior" of a pseudo-caustic was noted first in 
an analysis of the isothermal sphere of an infinite extension, whose 
projected mass density is inversely proportional to the radius
(referred to as SIS: singular isothermal sphere), and
was attributed to the density singularity at the center \citep{kovner87}.
The nature of the pseudo-caustic can be readily read off from the lens 
equation. 
\opeqn
 \omega = z - \dfrac{\eta}{\bar z}
 \label{eqSIS}
\cleqn
where $\eta$ is the radial parameter of $z$: $z = \eta a e^{i\theta}$
and $a$ is a reference radius the mass within which is used to 
normalize the lens equation. Eq.(\ref{eqSIS}) can be obtained 
by using the Newton's theorem of a spherical symmetric mass or 
by calculating the deflection angle as in the Appendix.

If we consider a monopole lens, for which $\eta$ should be replaced 
by $1$, an $\epsilon$-circle around the singularity at the origin is 
mapped to a large circle at infinity and the lens equation is a
double-covering mapping. In other words, the image plane is 
divided into two regions by the critical curve, which is mapped to the origin
of the source plane, and each of the two regions is mapped to
the entire source plane. 

Now $z=0$ is a singularity of eq.(\ref{eqSIS}) 
where infinitely many finite-size vectors clash. Speaking in terms 
of the vector field considered by \citet{burke81}, the vector field
is not well defined at $z=0$. If we cut out an
$\epsilon$-hole, the vector field behaves well in the neighborhood
of the punctured origin, and the $\epsilon$-boundary around the singularity 
is mapped to a circle of radius $1/a$ in the source plane. The $z$-plane
outside the critical curve covers the whole $\omega$-plane, but the 
punctured disk inside the critical curve covers only the disk of radius 
$1/a$ centered
at the origin of the $\omega$-plane. In other words, the disk of radius 
$1/a$ of the $\omega$-plane has two images and and the rest of the
$\omega$-plane has only one image; the number of images changes
as the source crosses the circle of radius $1/a$ or the pseudo-caustic.
As the source crosses from inside the disk of radius $1/a$, one of the two 
images ``disappears into the $\epsilon$-hole."  The disappearing image
is the negative image because the $\epsilon$-hole is inside the critical
curve. In summary, the pseudo-caustic is
an indicator that the lensing equation is not a full double covering
but a ``one and half" covering of the source plane and 
that the total parity of the images is not an invariant. 

If we truncate the isothermal sphere cylindrically such that the truncated 
mass is infinitely elongated along the line of sight, the projected mass 
density  is inversely proportional to the radius, and the inside lens equation 
is given by the same equation as eq.(\ref{eqSIS}) where $a$ is the radius
of the mass disk. The reason why the lens equations are the same is 
that only the mass inside the probing circle affects the two-dimensional
gravitational field at the probing point $z$ 
as is well known as the Newton's theorem. The deflection
angle is essentially, apart from some constants including the distance 
factor, the two-dimensional gravitational field.

\subsection{Constant Density ESL Inside Equation}

The deflection angle of the ESL
can be expressed as an elementary function 
when the density is constant  \citep{FKK07}. 
The (normalized) lens equation inside the mass is a linear equation
\opeqn
 \omega = \left(1-\dfrac{1}{ab}\right) z 
   + \left(\dfrac{1}{ab}\dfrac{a-b}{a+b} \right) \bar z
\cleqn
where $a$ and $b$ are the semi-major and semi-minor axes of the mass
ellipse. The lens equation is well defined
everywhere., and there is no pseudo-caustic.
The equation has always one solution and needs to be screened 
by hand or by studying the mass boundary and its secondary curves
(map the mass boundary 
by the lens equation into a curve in the source plane and study the
pre-image curves in the image plane that are mapped to the curve
in the source plane; one of the pre-image curves is the mass
boundary)
if the solution is inside the mass and hence is an actual image. 
$\omega=0$ has a solution at $z=0$, and $\omega=\infty$ has a
solution at $z=\infty$. Therefore a sufficiently large $\omega$
will fail to form an image inside the mass where the equation is
valid. 

The same equation applies to an infinite mass constant density ESL with
an infinite extension, which is due to the well known Newton's theorem 
for the gravitational field of an elliptically symmetric mass.

\subsection{Arcsine Lens Inside Equation}

The deflection angle of the ESL
can be expressed as an arcsine function 
when the projected mass density is inversely proportional to the 
elliptic radius \citep{KL09}. The mass model is 
commonly referred to as a SIE (singular isothermal ellipsoid)
since it was named as such in 
\citet{kormann} 
even though it was neither proved nor demonstrated to be an isothermal
state . We shall call it arcsine lens.
The inside lens equation of an arcsine lens is given
as follows as can be found in the Appendix.
\opeqn
  \omega = z - \dfrac{1}{c} \arcsin \left(\dfrac{c\eta}{\bar z}\right) 
  \label{eqArcsinInside}
\cleqn
where $c$ is the focal length and $\eta$ is the elliptic radial parameter of $z$.

As in the case of the SIS, the elliptic version eq.(\ref{eqArcsinInside})
is ill-defined at $z=0$, and an $\epsilon$-circle around the singularity
is mapped to a finite loop in the source plane as $\epsilon \rightarrow 0$. 
The arcsine lens inside equation has a pseudo-caustic. The curve is 
a simple loop given by 
\opeqn
 w = -\dfrac{1}{c} \left(\dfrac{c}{a\cos\theta- i b\sin\theta}\right) \ .
\cleqn
The pseudo-caustic loop has the same sense with the $\epsilon$-circle
but has its phase lagged by $\pi$.

\citet{kormann} called the pseudo-caustic curve a cut following the
convention used by \citet{kovner87} and observed that the pseudo-caustic
can intersect with the caustic curve. The caustic is a symmetric astroid
and is either enclosed inside or intersect with the pseudo-caustic curve.

In order to calculate the critical curve and caustic curve, it is convenient
to use the elliptic coordinates:  $z=\eta(\cos\theta+i b \sin\theta)$ 
because the deflection angle is independent of the elliptic radius.
The partial derivative with respect to $z$, $\partial_z$, 
can be expressed in terms of the partial derivatives in elliptic coordinates,
\opeqn
 \partial \equiv \partial_z 
   = \dfrac{a\sin\theta+ib\cos\theta}{2 iab} \partial_\eta
     + \dfrac{a\cos\theta-ib\cos\theta}{2\eta iab} \partial_\theta
 \ , 
\cleqn
and the Jacobiant matrix 
components of the inside lens equation (\ref{eqArcsinInside}) are 
\opeqn
 \partial\omega = 1 - \dfrac{1}{2\eta a b}\ ; \qquad
  \bar\partial\omega = \dfrac{1}{2\eta a b}\dfrac{z}{\bar z} 
\cleqn
where $\bar\partial$ is the complex conjugate of $\partial$.
The Jacobian determinant is  
\opeqn
  J = |\partial\omega|^2-|\bar\partial\omega|^2 = 1 - \dfrac{1}{\eta a b}
   \  , 
\cleqn
and on the critical curve
\opeqn
 \eta = \dfrac{1}{a b} \ .
\cleqn 
The critical curve is an ellipse.
\opeqn
 z= \dfrac{1}{ab}(a\cos\theta + i b \sin\theta) \ .
\cleqn

The corresponding caustic is a quadroid (or astroid) 
with the four cusps on the 
real and imaginary axes (or symmetry axes). The precusps can be found
using the same method as in  \citet{RCB}. 
Since $\partial\omega$ is
real, the precusp condition $0 = \partial_- J$ becomes 
\opeqn
  0 = z \partial J - \bar z \bar\partial J \ .
\cleqn 
The equation leads to $\sin 2\theta = 0$, hence the precusps are at
$\theta = 0$, $\pi/2$, $\pi$, and $3\pi/2$. The precusps are mapped to
cusps on the positive real, negative imaginary, negative real, 
and positive imaginary axis respectively. 
\opeqn
 \Delta \omega_{real} \equiv \omega_0 - \omega_{\pi}
        = \dfrac{2}{b} - \dfrac{2}{c}\arcsin (\dfrac{c}{a})
\cleqn
\opeqn
 \Delta\omega_{imag} \equiv \omega_{\pi/2} - \omega_{3\pi/2}
        = \dfrac{2i}{a} - \dfrac{2}{c}\arcsin (\dfrac{ic}{b})
\cleqn
$\Delta\omega_{real} > 0$ and $\Im\Delta\omega_{imag} < 0$,
and hence the caustic curve has the opposite sense with the critical
curve.

\begin{figure}
\begin{center}
\includegraphics[angle=0,scale=0.45,keepaspectratio]{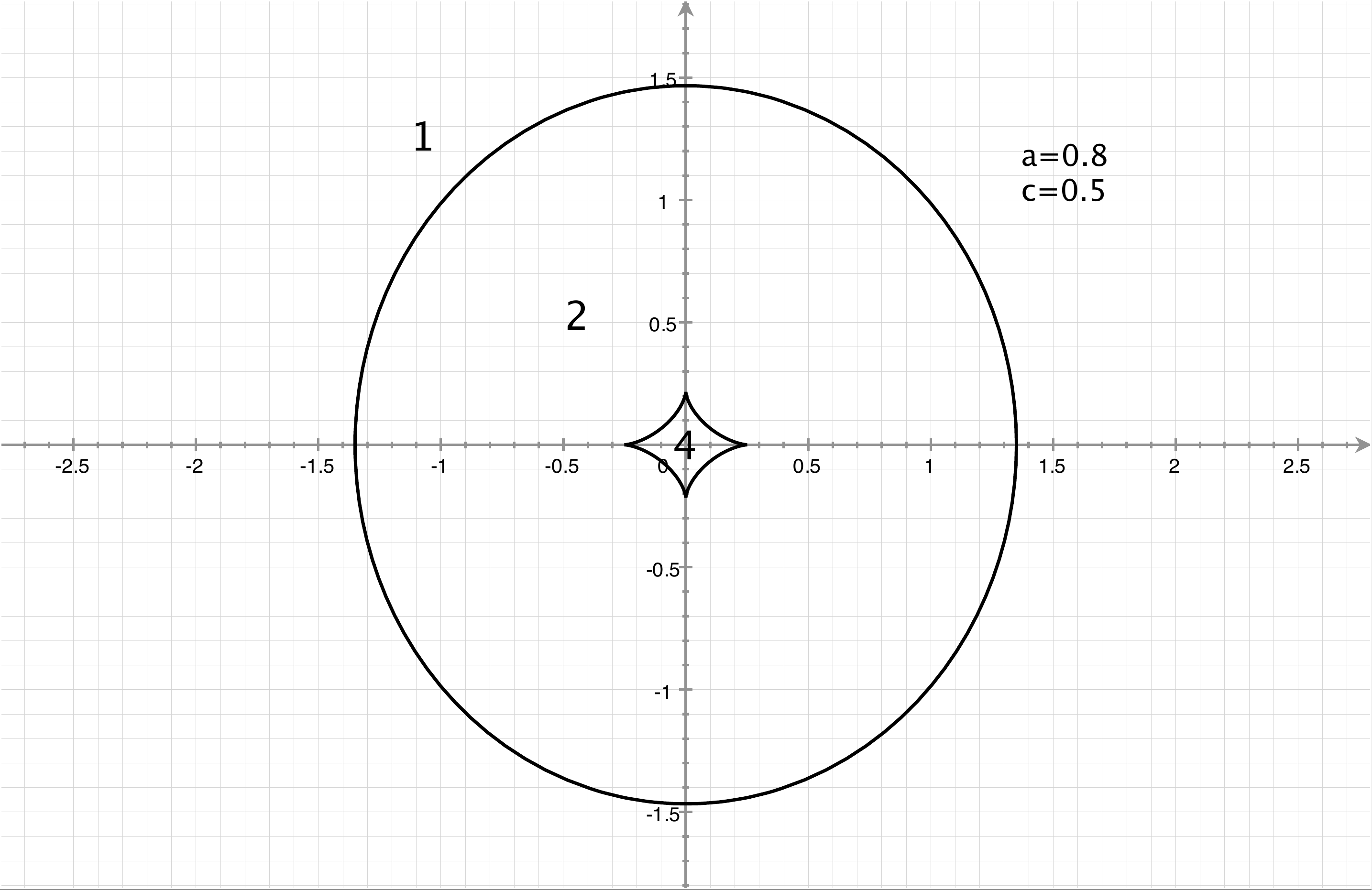}
\includegraphics[angle=0,scale=0.45,keepaspectratio]{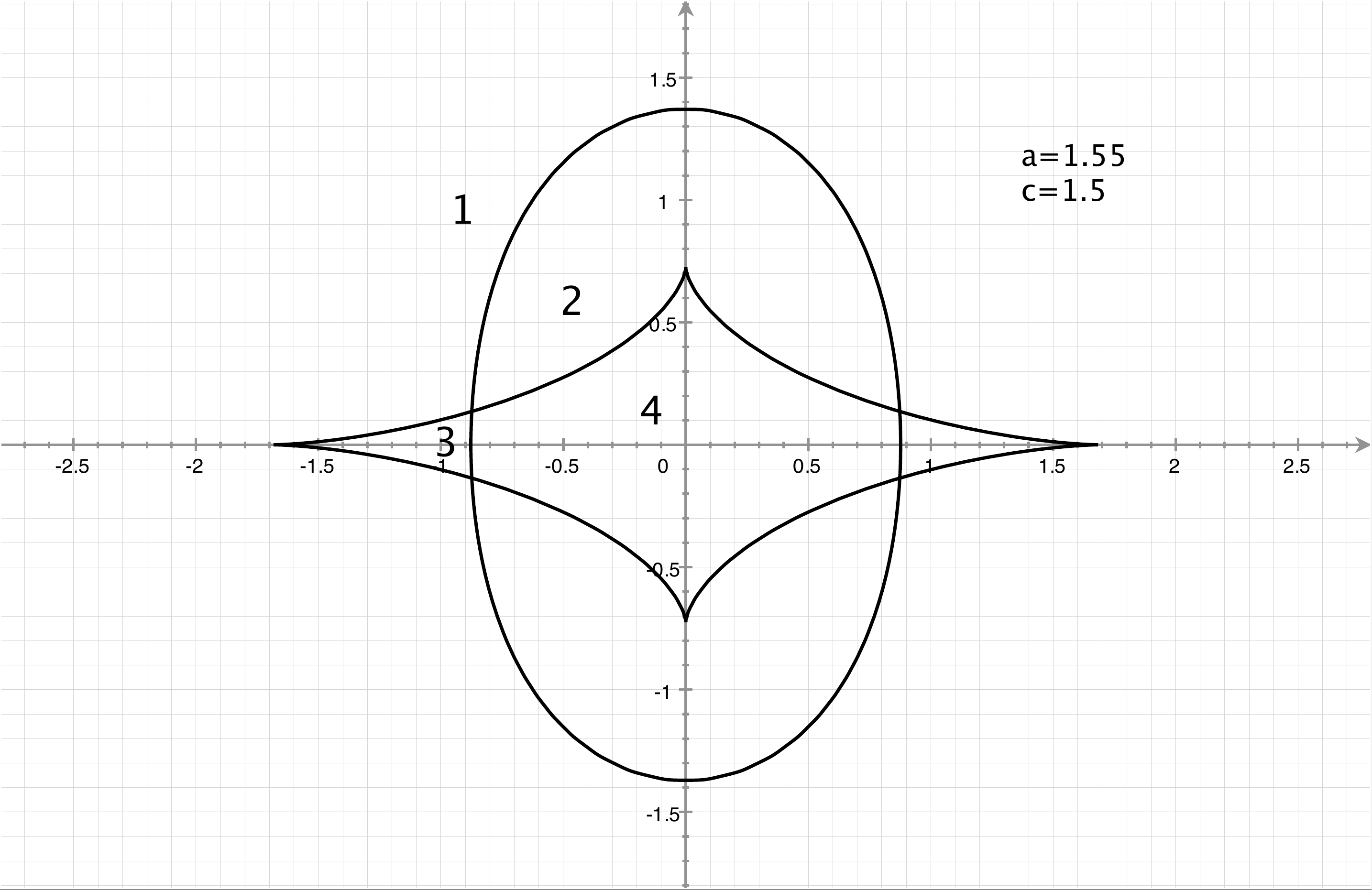}
\end{center}
\caption{There are two morphologies of the caustic domains of the Arcsin lens inside equation: 
1) the caustic is enclosed inside the pseudo-caustic: 
the images are of
$1/0$, $1/1$, and $2/2$ where $m$ and $n$ of $m/n$ are the numbers 
of positive and negative images.
2) the caustic and pseudo-caustic intersect: the images are of
1/0, 1/1, 2/1, and 2/2.The caustics are cuspy and the pseudo-caustics
are smooth. The numbers in the caustic domains indicate the total
number of images of each caustic domain. The total parity is 0 or1
for both cases.
} 
\label{fig-ICSO}
\end{figure}

Figure \ref{fig-ICSO} shows the cases of the caustic and pseudo-caustic. 
The number of the images of the caustic zones are indicated and clearly
demonstrate that the total parity of the images is not an invariant.
There is no case where the pseudo-caustic is enclosed inside the caustic 
curve because the cusps
on the imaginary axes are always inside the pseudo-caustic. The points of
the pseudo-caustic and caustic on the positive imaginary axis 
(semi-minor axis of the mass ellipse) are from
$\theta = 3\pi/2$, and it is easy to see that 
$\Im(\omega_{pseudo}(3\pi/2) - \omega_{caustic} (3\pi/2)) = 1/a > 0$.

\subsection{Truncated Constant Density ESL Outside Equation}

We have discussed a few cases in which a pseudo-caustic arises
because of a point singularity where Burke's vector field is not well
defined and around where the vector values are not all infinite. 
In this section
we will introduce a line singularity where Burke's vector field is not 
well defined and around where the vector values are not all infinite. 

Consider truncating the constant density ESL at $t=1$, and the 
lens equation valid outside the mass is given by
\opeqn
 \omega = z - \dfrac{2}{c^2} (\bar z - \sqrt{\bar z^2 - c^2}) \ .
\label{eqConstOutside}
\cleqn
$z = \pm c$ are the branch points of the square root function and
we need to place branch cuts to establish a convention where the
Riemann sheets are sewed up. 
Mathematically a branch cut can be placed wherever we 
are pleased to put it as far as it is connected to the branch
point in concern. 
Physically, however, we are constrained because two images
in an $\epsilon$-neighborhood cannot come from two widely 
separated sources unless there is a physical reason. Placing the
branch cuts randomly will induce such an unphysical behavior.
The outside lens equation (\ref{eqConstOutside})
depends only on the focal length $c$ and is valid for an infinite 
family of confocal lenses for which the image zone can be
everywhere except for an $\epsilon$-ellipse around the line
segment connecting the two focal points or the two branch points. 
Thus physically the line segment connecting the two focal points
is the only option as the branch cut. The square root function 
should be understood as 
$\sqrt{z^2 - c^2}=(r_+ r_-)^{1/2} e^{i(\theta_+ + \theta_-)/2}$
where $z \pm c = r_\pm e^{i\theta_\pm}$. 

Across the branch cut, the (complex) deflection angle changes the
phase by $\pi$. In other words, 
Burke's vector field is not well defined on the
branch cut because it is double-valued. We can cut along the
branch cut to remedy it. The boundary of the cut is mapped to 
a finite loop by the lens equation. The finite loop is the 
pseudo-caustic, and across which a negative image
disappears into or appears from the branch cut.

\subsection{Arcsine Lens Outside Equation}

If we truncate the arcsine lens at $t=1$, the outside lens equation
is given by
\opeqn
 \omega = z - \dfrac{1}{c} \arcsin \left(\dfrac{c}{\bar z}\right) \ .
 \label{eqArcsinOutside}
\cleqn
The deflection angle is double-valued on the line segment
connecting the two focal points. Call it c-cut.  Cut the
image plane along the c-cut, and the boundary of the c-cut
is mapped to a semi-infinite loop whose extension is finite
in the direction of the c-cut. The semi-infinite loop is the 
pseudo-caustic. 

\citet{BE10} (BE10 herefrom) studied the number of images 
of the lens equation (\ref{eqArcsinOutside}) using a converted 
equation and found that the number of images can be anywhere 
from 1 to 6. Since the functional BE10 used is different from
that of (\ref{eqArcsinOutside}), the caustic planes in BE10 
should be reinterpreted. It turns out that the transformation
is simple: $\omega^\prime = - c \omega$ where $\omega^\prime$
is the source plane variable of the associated equation, and 
the morphology of its caustic plane can be read almost as if it
were the caustic plane of the original equation 
(\ref{eqArcsinOutside}) if one overlooks the absolute value of 
the coordinates and takes into account of the reversing of 
the parities. The pseudo-caustic in a caustic plane in BE10
arises from the image plane boundary (thus the converted 
equation cannot be the lens equation of 
a realistic lens), and it corresponds to the pseudo-caustic
that arises from the c-cut that is essentially from a branch
cut. 

\begin{figure}
\begin{center}
\includegraphics[angle=0,scale=0.45,keepaspectratio]{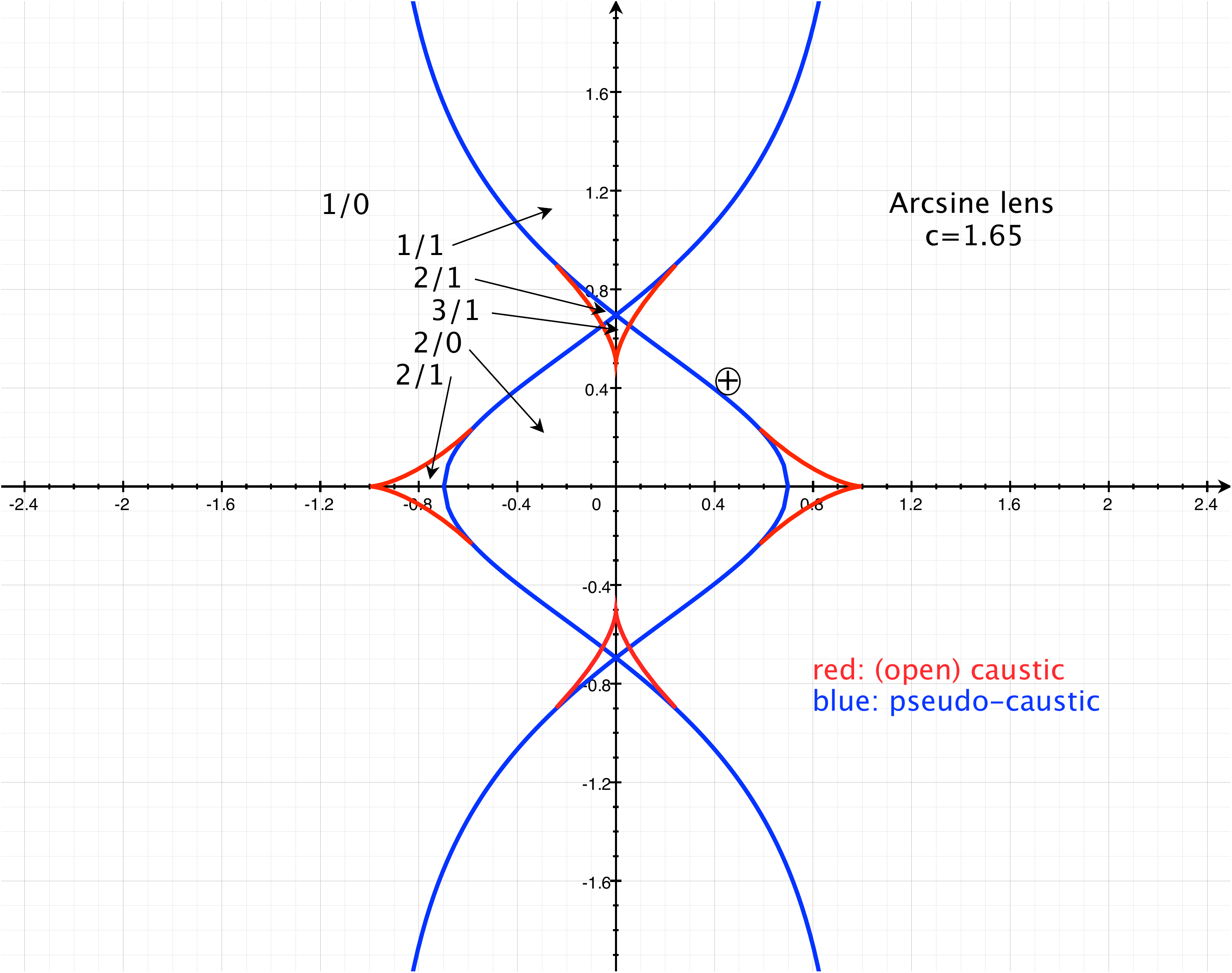}
\end{center}
\caption{The caustic domains of the arcsine lens outside equation
with $c=1.65$.
The red is the (open) caustic and the blue is the pseudo-caustic.
At the joining points, the caustic and pseudo-caustic are tangential.
$m/n$ stands for \#(positive images)/\#(negative images). The total 
parity is 0, 1, or 2. $\oplus$ indicates that the image crossing the 
segment of the pseudo-caustic and its symmetric counterparts
is a positive image. For the rest of the pseudo-caustic, the image
appearing or disappearing is a negative image.
} 
\label{fig-asinCD}
\end{figure}

Figure \ref{fig-asinCD} shows the caustic domains for $c=1.65$.
The critical curve bifurcates at $c=\sqrt{2}$, 
and at $c=1.65$ it is
made of three loops which enclose the origin
and two focal points respectively. The caustic curves
are open and closed by the pseudo-caustic curve. 
The transition from the domain 1/0  to the domain 2/0 is 
by an increase of a positive image as is indicated  
in the notations, and the reason for the relevance of
a positive image is that the segment of the c-cut
is outside the critical curves. One of the four ``positive" segments
of the pseudo-caustic is indicated by the symbol $\oplus$.

\subsection{Arbitrary Bounded Mass Lenses}

We have examined the pseudo-caustics of a few
algebraic lens equations. For an arbitrary elliptically symmetric 
lens, the equation is an integro-algebraic equation. However 
the existence or non-existence of the pseudo-caustic arising from
the c-cut can be easily discerned by looking at the behavior of the
c-cut at the origin, $z=0$. Since the deflection angle is continuous
on the $\epsilon$-boundary of the c-cut, double-valuedness at 
$z=$ is an indicator that the c-cut is mapped to a finite loop, {\it i.e.,}
a pseudo-caustic.
The ESL outside lens equation is in general, 
as can be found in the Appendix,
\opeqn
 \omega = z - \dint_0^1 \dfrac{\tilde\sigma(t) t dt}{(\bar z^2-c^2t^2)^{1/2}} 
\cleqn
where $c$ is the focal length of the mass boundary ellipse. 
At $z=0\pm i0$,
\opeqn
     DAngle = \int_0^1 \dfrac{\pm i}{c} \tilde\sigma(t) dt \ .
\cleqn
It is double-valued and finite.
At the intersections of the $\epsilon$-boundary
of the branch cut and the real line, $z = \pm (a\eta + i0)$, the deflection 
angle is single-valued and finite. Therefore the peudo-caustic is a finite loop.
 
The ESL inside lens equation, also found in the Appendix, is
\opeqn
 \omega = z - \dint_0^\eta \dfrac{\tilde\sigma(t) t dt}{(\bar z^2-c^2t^2)^{1/2}} 
\cleqn
where $c$ is the focal length of the mass boundary ellipse
and $\eta$ is elliptic radial parameter of $z$. 
$z=0\pm i0$ are mapped to the single value $DAngle=0$ because the 
integral upper bound $\eta = 0$. Any other point $z$ on the 
 branch cut is also single-valued because it is always outside
 the branch cut connecting the branch points $\pm c\eta$.
 Thus the ESL inside lens equation of an arbitrary bounded mass
 has no pseudo-caustic.
 
 The examples of the bounded ESL lenses discussed in the previous
 subsections have sharp cutoffs at $t=1$. However, the class of bounded 
 density functions include smooth functions. For example, 
 a family of bounded smooth
 functions can be constructed as follows \citep{eremenko}.
 \begin{eqnarray}
    \sigma(t) \propto 
       &  e^{-\alpha^2/(1-t)^2}:   & \qquad t < 1 \\
       & 0:  &  \qquad  t \geq 1
\end{eqnarray}
Thus smooth bounded density functions are not immune from pseudo-caustics
and 
contradict the smooth density function arguments for the Burke's theorem.

\section{Summary and Discussion}
\label{secThree}

Burke's theorem invokes Poincar\'e-Hopf index theorem of the
vector field on the 
compactified complex plane as a sphere. The index of a loop 
is determined
only by the zeros of the vector field enclosed in the loop. 
The behavior of 
Burke's vector field is determined by the behavior of the deflection
angle where the latter is an integral of a mass density function with
the kernel $(\bar z - \bar \xi)^{-1}$ where $z$ is the probing position and
$\xi$ is the position variable of the density function. If the vector
field is continuous everywhere, then the index of the total number of
the images can be accounted for by drawing a large loop at infinity. 
The large
loop at infinity encloses one positive (unmagnified) image at infinity 
when the source is at infinity, and hence the index of the large loop 
is 1 and Burke's theorem follows.

However, the deflection
angle integral is not necessarily continuous. 
The best known singularity is that
of a monopole or a monopole-equivalent (due to the Newton's 
theorem), and it introduces a hole in the sphere of the complex
plane. In other 
words, the underlying manifold is not compact, and strictly, the
Poincar\'e-Hopf index theorem does not apply. However, a pole
respects the total parity invariance, and the hole due to the 
monopole singularity only violates the odd-numberedness 
of the Burke's theorem; the 
asymptotic number of the images is determined by the zero at
infinity and the number of the poles due to the monopole lens 
positions;  the total parity of the $n$-point lens is $1-n$. 

The cylindrically
truncated isothermal lens is another example of the singularities
and has been known for causing a pseudo-caustic violating the
total parity invariance. The mass density singularity at the origin
is responsible for the pseudo-caustic, and it has been argued 
that Burke's theorem should hold if the mass density function is
smooth. 

Here we have shown another type of singularity that is
common in bounded elliptically symmetric lenses; the new type
of singularity comes from a branch cut and a smooth ESL mass 
density function is not immune from it.  In conclusion, a smooth
mass density function is not a guarantee for the validity of the
Burke's theorem. 

Morse theory studies \citep{mackenzie,petters} had not turned up
the branch cut singualrities that we have found in the bounded
elliptically symmetric lenses that are physically perfectly reasonable. 
It may be the time to reexamine 
the Morse theory in relation to gravitational lensing 
or null geodesics.

Point masses are the foundation of the Galactic 
microlensing experiments
including searches for microlensing planets. The point masses
preserve the total parity invariance but do not respect the odd
number theorem because an odd number of point masses 
produces an even number of images asymptotically. Here we
found a line singularity of bounded elliptic masses that is 
related to a branch cut and causes a pseudo-caustic. 
A pseudo-caustic violates the total 
parity invariance.
A bounded elliptically symmetric mass is a reasonable model
for galaxy lenses or cluster lenses.  
Are there other types of singularities
yet to be discovered in gravitational lensing?

One can consider a smooth density function that is unbounded
but asymptotically falls reasonably fast to avoid a
pseudo-caustic and an infinite mass. For example, a bounded
mass density function that reasonably fits a real lens may be 
reasonably approximated by 
truncating the large frequency modes of its Fourier integral.
The truncated Fourier integral should be infinitely differentiable
and infinitely extended. Then does the real lens respect or 
violate the invariance of the total parity? If we insist on the
preservation of the total parity invariance, we need to assume
that the lens mass is infinitely extended. However small a
density it might be at a large distance, we need to assume
the elliptic symmetry. Is it sensible to do so? 
One practical approach may be to investigate
the behavior of the pseudo-caustics of bounded masses that 
are close to or intersect with their caustics in the process 
of increasing the differentiability. It will be explored in a separate
work.

\appendix

\section{SIS or Cylindrically Truncated Isothermal Lens}

If we consider a self-gravitating isothermal sphere of gas mass in 
thermal equilibrium,
the algebraically known solution of the Lane-Emden equation,
the so-called SIS (singular isothermal sphere), has the density profile
that is proportional to the inverse radius square. 
The total mass is infinite because of the indefinite extension of the
constant mass shells. 
In practice, the tidal interaction with the neighboring systems are
likely to define the edges and mass breaks of the self-gravitating
objects. If we truncate the SIS at a certain (3-d) radius, say $r=a$,
and integrate for the projected mass of the lens, the 2-d mass density
function has the form
\opeqn
 \Sigma(t) \propto \frac{1}{t}\arctan\left(\dfrac{(1-t^2)^{1/2}}{t}\right)
  \ ; \qquad
  t \leq 1
\label{eqSigmaCirc}
\cleqn
where the position variable is expressed in terms of the circular coordinates
as $\xi = t a(\cos\theta + i \sin\theta)$ and the mass trucation is made at
$t=1$. In order to find the lens equation, we can use the Newton's theorem
that the gravitational field of a circularly symmetric mass is determined by
the mass inside the circle of the radius  of the probing point placed at
the center, and hence we need to integrate the density function in 
eq. (\ref{eqSigmaCirc}) from the center to an arbitrary $t$ less than 1.
The integration doesn't relent to
a nice manageable algebraic function. So we cheat a bit (maybe a lot)
as is customary and ignore the arctangent factor to obtain an easy-to-handle
density function $\Sigma(t) \propto 1/t$ and truncate it at $t=1$. In other 
words, the mass being considered is the infinite isothermal sphere cut into
an infinite cylinder that is infinitely long along the line of sight. 
Thus the (2-d) mass distribution may be best referred to 
as a cylindrically truncated isothermal lens (CTIL). 

The gravitational lens equation is given by
\opeqn
   \omega = z - 4GD \dint \frac{\Sigma (t) }{\bar z - \bar \xi}\, d^2\xi \ 
\label{eqLeq}
\cleqn
$\omega$ and $z$ are variables for a source and its image in the lens
plane at the distance of the lens and 
$D$ is the reduced distance: $1/D = 1/D_1 + D_2$ where
$D_1$ and $D_2$ are the distances from the lens to the observer and the source.
The area element $d^2\xi = d\xi_1 d\xi_2 = (-2i)^{-1} d\xi d\bar\xi$.

If the mass of the lens is $M$, introduce the normalized density
function $\sigma(t) = \Sigma(t)/M$
and renormalize the position variables by the Einstein ring radius of the
mass $M$, $R_E = (4GMD)^{1/2}$, so that the unit distance of the lens plane
is given by $R_E$. The lens equation (\ref{eqLeq}) is rewritten as
\opeqn
 \omega = z - \dint \frac{\sigma (t) }{\bar z - \bar \xi}\, d^2\xi \ .
\label{eqLeqNormalized}
\cleqn 

Since $\sigma(t)$ depends only on $t$, we can manipulate the 
deflection angle integral as follows. 
If $\partial D_t$ denotes the circle of radius $ta$ and
$D_t$ is the circular disk inside the boundary circle 
$\partial D_t$, then the area of the annulus $D_{t+dt} -D_{t}$ is
of the order of $dt$ and $\sigma(t)$ can be considered constant in the annulus.
\opeqn
 \left[ \int_{D_{t+dt}} - \int_{D_t} \right]
  \left[\dfrac{\sigma(t)d^2\xi}{(\bar z - \bar \xi)}\right]
  =\sigma(t) \left[ \int_{D_{t+dt}} - \int_{D_t} \right]
  \left[\dfrac{d^2\xi}{\bar z - \bar \xi}\right]
  = \sigma(t) dt \dfrac{\partial}{\partial t}
    \left[\int_{D_t} \dfrac{d^2\xi}{\bar z - \bar \xi}\right] \ .
\cleqn  
Therefore the deflection angle DAngle is give by 
\opeqn
 DAngle = \int_0^1 dt \, \sigma(t) \dfrac{\partial}{\partial t}
    \left[\int_{D_t} \dfrac{d^2\xi}{\bar z - \bar \xi}\right] \ .
\label{eqDAngle}
\cleqn

We can calculate the integral inside the square bracket 
in eq(\ref{eqDAngle}), which we shall label $D_t$Integral,
using the Schwarz function of
the circle and three facts are useful to know:
the Stoke's theorem or Green's theorem in two dimensions, 
the Cauchy integral of an analytic function, 
and the Schwarz function of the circle. 
In this case of the circular mass boundary, it is simpler to 
apply the Newton's theorem, but we will develop the
machinery that can be also used for an elliptic mass
boundary.

The Stoke's theorem is well known in arbitrary dimensions because of their
physical relevance in electrodynamics, fluid dynamics, etc. See for example 
\citet{jackson}. In two dimensions,
\opeqn
  \int_{\partial \Omega} u dx + v dy
    = \int_\Omega (\partial_x v - \partial_y u) dxdy \ ,
\cleqn
in which it is implicit that $u$ and $v$ are differentiable and
the differentials are continuous, and which 
can be rewritten in terms of the exterior differentiation $d$.
\opeqn
  \int_{\partial \Omega} u dx + v dy
    = \int_\Omega d(u dx + v dy) 
\label{eqGreen}
\cleqn
(Sometimes $d\wedge$ is used instead of $d$ in order to make the
exterior differentiation more explicit.)   
If $u$ and $v$ are $x$ and $y$ components of vector $\vec w$,
it can be written in the vector form.
\opeqn
  \int_{\partial \Omega} \vec w \cdot d\vec l
    = \int_\Omega \nabla\times \vec w \, dx dy
    \ .
\cleqn
The real space Stoke's theorem can be readily 
translated into the complex form in the complex plane using
eq.(\ref{eqGreen}), and it states \citep{gong} 
that if $w = w_1 dz + w_2 d\bar z$ is an exterior differential one
form in a domain $\Omega$ and $\partial\Omega$ is the boundary of $\Omega$,
where $w_1(z,\bar z)$ and $w_2(z,\bar z)$ are differentiable once and 
the differentials are continuous,
then
\opeqn
  \int_{\partial\Omega} w = \int_{\Omega} dw
\cleqn  
(The orientation of the boundary curve $\partial\Omega$ is assumed to be
counterclockwise as usual. If one walks along the boundary, the left hand
points to the domain $\Omega$. Walking along a loop with the right hand in 
is the negtaive of the left hand in.)

The Cauchy integral of $f(\xi)$ is defined as \citep{titch}
\opeqn
   \dfrac{1}{2\pi i} \int_\Gamma \frac{f(\xi)}{\xi - z}\, d\xi
\cleqn
where $f(\xi)$ is analytic inside and on a simple closed curve $\Gamma$. 
The integral is equal to $f(z)$ if $z$ is inside $\Gamma$ and the 
equality is called Cauchy's integral formula; 
it is equal to $0$ if $z$ is outside $\Gamma$ because the integrand 
is analytic (Cauchy's theorem). The derivative of $f(z)$ 
can be expressed in terms of the integral of $f(\xi)$ and a kernel.
\opeqn
  f^\prime(z) = \dfrac{1}{2\pi i} 
    \int_\Gamma \frac{f(\xi)}{(\xi - z)^2}\, d\xi
\cleqn

The Schwarz reflection symmetry principle \citep{needham}
states that given a smooth enough 
arc in the complex plane, there is an analytic function (in a domain that
includes the arc) that maps the arc into its complex conjugate. The analytic
function is called the Schwarz function of the arc and commonly denoted
$S(\xi)$. The complex conjugate $\overline {S(\xi)}$ reflects the domain with 
respect to the arc. On the arc, $\xi = \overline {S(\xi)}$, or $\bar \xi = S(\xi)$.
Here the relevance is that the Schwarz function of the circle is nothing
but $S(\xi)=t^2 a^2/z$ because on the circle $\xi \bar \xi = t^2 a^2$
where a position variable in the circular coordinate is written as
$\xi = t a (\cos\theta + i \sin\theta)$.

Now we are ready to integrate the deflection angle over the unit
circular disk $D$.

\subsection{CTIL Outside Equation}

If the probing point $z$ is outside the mass disk, the integrand
of the $D_t$Integral does not have a singularity and is analytic.
The complex conjugate of the $D_t$Integral is
\opeqn
 \overline{D_tIntegral} = \int_{D_t} \frac{d^2\xi }{z - \xi}
  = \frac{1}{2i}\int_{D_t}\frac{d\bar\xi d\xi }{z - \xi}\, 
  =\frac{1}{2i}\int_{D_t} d\left(\frac {\bar\xi d\xi}{z - \xi}  \right)
\cleqn
\opeqn
 = \frac{1}{2i}\int_{\partial D_t} \dfrac {\bar\xi d\xi}{z - \xi} 
 =  \frac{1}{2i}\int_{\partial D_t} \dfrac {t^2 a^2 d\xi}{\xi(z - \xi)} 
 = \frac{\pi t^2 a^2}{z}
\cleqn
where $z$ is outside the boundary $\partial D_t$.
Therefore the deflection angle in eq.(\ref{eqDAngle}) is
\opeqn
 DAngle = \int_0^1 dt \sigma(t) 
  \frac{\partial}{\partial t} \left(\frac{\pi t^2 a^2}{\bar z}\right)
                = \int_0^1 \frac{2\pi a^2\, \sigma(t) t dt}{\bar z}
                =  \int_0^1 \frac{\tilde\sigma(t) t dt}{\bar z}
\cleqn
where
\opeqn
 1 = \int_0^1 \tilde\sigma(t) t\, dt \ ,
\cleqn
and hence the outside lens equation is given by the 
monopole lens equation.
\opeqn
 \omega = z - \frac{1}{\bar z} \ .
\cleqn
The outside lens equation of a circularly symmetric lens
is independent of the mass density function.

\subsection{CTIL Inside Equation}

If the probing point $z$ is inside the mass disk $D$, 
we need to mind the singularity at $\bar \xi = \bar z$ 
in $D_t$Integral. 
If we cut out from the mass disk $D$ 
a small disk of radius $\epsilon$ centered at $z$, $D(z,\epsilon)$,
then we can apply the Green's theorem to the $D_t$Integral over 
$D_t - D(z,\epsilon)$.  
\opeqn
  \dint_{D_t- D(z,\epsilon)} \dfrac{d^2\xi}{z-\xi}
  = \dfrac{1}{2i}
    \dint_{\partial D- \partial D(z,\epsilon)} 
       \dfrac{\bar \xi d\xi}{z - \xi} \ .
\label{eqGreenInside}
\cleqn
Now the surface integral over $D(z,\epsilon)$ on the LHS vanishes,
\opeqn
 \dint_{D(z,\epsilon)}  \dfrac{d^2\xi}{z-\xi}
  = \int_0^{2\pi} \int_0^{\epsilon}\dfrac{rdr d\theta}{r e^{i\theta}}
  = \epsilon \int_0^{2\pi} \dfrac{d\theta}{e^{i\theta}} = 0 \ ,
\cleqn
hence the $D_t$Integral is given by the RHS of eq.(\ref{eqGreenInside}).
\opeqn
 \overline{D_t Integral}
  = \dfrac{1}{2i}
    \left[\dint_{\partial D_t} - \dint_{\partial D(z,\epsilon)}
     \right]
       \dfrac{\bar \xi d\xi}{z - \xi}  \ .
\cleqn
The first contour integral over $\partial D_t$ vanishes if $z$ is
inside $D_t$ because the integrand is analytic outside the
contour; if $z$ is outside the contour, the integral picks up the
contribution from the pole at $\xi=0$ inside the contour or 
the pole $\xi=z$ outside the contour which result in the same
value.
\opeqn
   \dfrac{1}{2i}
    \int_{\partial D_t}        \dfrac{\bar \xi d\xi}{z - \xi} 
   =\frac{1}{2i} \int_{\partial D_t}       \dfrac{t^2 a^2 d\xi}{\xi(z - \xi)} 
     = \pi \frac{t^2 a^2}{z} \quad : \qquad \eta \geq t \ .
\cleqn
where $z = \eta a (\cos\theta + i\sin\theta)$.
The second contour integral over $\partial D(z,\epsilon)$ can be 
calculated using the Schwarz function of the circle:
$(\bar \xi- \bar z)(\xi - z) = \epsilon^2$.
\opeqn
   \dfrac{1}{2i}
    \dint_{\partial D(z,\epsilon)}
            \dfrac{\bar \xi d\xi}{z - \xi} 
   = \frac{1}{2i} \dint_{\partial D(z,\epsilon)} 
     \left( \dfrac{\bar z }{z-\xi} 
        - \dfrac{\epsilon^2}{(\xi-z)^2}\right) d\xi
   = - \pi \bar z \quad : \qquad \eta < t
\label{eqContourHole}
\cleqn
The contribution comes from the first term only. The second 
term vanishes because the derivative of the analytic function
$\epsilon^2$ (constant) is zero. Since the result is independent
of $t$, it does not contribute to the deflection angle integral.
Thus the deflection angle at $z$ inside the mass disk is given as follows.
\opeqn
 DAngle = \int_0^\eta \dfrac{\tilde\sigma(t) t\, dt}{\bar z} \ .
\cleqn
For a CTIL, $\tilde\sigma(t) = 1/t$, and the inside lens equation 
is given by
\opeqn
\omega = z - \dfrac{\eta}{\bar z} \ .
\cleqn

\section{Elliptically Symmetric Lenses}

For galaxy lenses that have elliptic shapes and light distributions, 
one can consider a class
of mass distributions whose projected densities depend on only the
elliptical radius. For example, $\Sigma(t) \propto 1/t$ where the 
position variable $\xi = t(a\sin\theta + i b\cos\theta)$.  
It is currently unknown if a singular elliptic mass distribution with
infinite or finite extension is exactly or closely related 
to a self-gravitating gas mass in thermal equilibrium.
However one can suspect that it might be the case 
at least for small ellipticities because of the existence of the circularly
symmetric system SIS and the fact that an angular momentum tends to add 
oblateness to the system.

Consider the family of ellipses parameterized by $t ( \geq 0)$ 
that fills the two-dimensional space.
\opeqn
  \xi = t (a \cos\theta + i b \sin\theta) \ , \qquad a \geq b
\cleqn
The curve $t = constant$ is an ellipse that is converted to the familiar 
equation for the ellipse in real coordinates $\xi_1$ and $\xi_2$ 
where $\xi = \xi_1 + i\xi_2$. 
\opeqn
  \frac{\xi_1^2}{a^2} + \frac{\xi_2^2}{b^2} = t^2 \ , \qquad t \geq 0 
\label{eqEllipse}
\cleqn
Assign a projected mass density profile truncated at $t=1$, 
and the deflection angle of an elliptically symmetric lens
(ESL) is given as follows similarly to eq.(\ref{eqDAngle}).
\opeqn
 DAngle = \int_0^1 dt \sigma(t) \dfrac{\partial}{\partial t}
  \left[\int_{\Omega_t} \dfrac{d^2\xi}{\bar z - \bar \xi} \right]
\label{eqDAngleEllipse}
\cleqn
where $\Omega_t$ is the ellipse with the elliptic radial 
parameter $t$.

In order to calculate the integral in the square bracket
of eq.(\ref{eqDAngleEllipse}), which we label $\Omega_t$Integral,
we need to calculate the
Schwarz function of the ellipse. 
Note that an analytic function is uniquely determined in a domain if it is
known in an arc in the domain. Thus it suffices to find the Schwarz function
on the ellipse where it is satisfied that $\bar z = S(z)$. Since we know the
equation of the ellipse in real coordiates, eq.(\ref{eqEllipse}), 
we only need to substitute $x = (z + \bar z)/2$ and $y = (z - \bar z)/(2i)$
and solve $\bar z$ in terms of $z$.
\opeqn
 \bar z = \dfrac{a^2+b^2}{c^2} z \pm \dfrac{2ab}{c^2}\left(z^2 - c^2 t^2 \right)^{1/2}  
\cleqn        
where $c^2 = a^2 - b^2$ and $z = \pm ct$ are the focii of the ellipse. 
The square root function is intrinsically a double-valued function and here 
one branch is chosen so that it is a univalent analytic function (except at
the branch points $z = \pm ct$). 
If $z \pm ct = r_\pm e^{i\theta_\pm}$, we choose 
$(z^2 - c^2 t^2 )^{1/2} = (r_+ r_-)^{1/2} e^{i(\theta_+ + \theta_-)/2}$.    
Then the Schwarz function of the ellipse is given by
\opeqn
 S(z) = \dfrac{a^2+b^2}{c^2} z - \dfrac{2ab}{c^2}\left(z^2 - c^2 t^2 \right)^{1/2}
  \ .
\cleqn
The other branch is not a solution because the function does not return $\bar z$.
(The readers are encouraged to check numerically.)

The Schwarz funciton can be decomposed into two parts: 
$S(z) = S_1(z) + S_2(z)$ where
\opeqn
S_1(z) = \dfrac{a-b}{a+b} z \quad ; \qquad
S_2(z) = \dfrac{2ab}{c^2} \left(z - (z^2 - c^2 t^2)^{1/2}\right)
\ .
\cleqn
$S_1(z)$ diverges at infinity and $S_2(z)$ is singular (branch points) at
the focii. Thus $S_1(z)$ is analytic inside the ellipse $\partial\Omega_t$ 
and $S_2(z)$ is analytic outside $\partial\Omega_t$.
(The decomposition is not essential for the integration even though one
can say it is convenient. Handling $S(z)$ directly without the decomposition 
by manipulating the pole at $z=\infty$ by an inverse transformation
is also instructive.)

Be reminded from the previous section that the singularity at $\xi=z$
when the probing point $z$ is inside the mass disk $D_t$ does not
contribute to the deflection angle. Thus, the only difference between
the outside lens equation and inside lens equation is the range of 
the integration in $t$ in eq.(\ref{eqDAngleEllipse}). Now assuming that
$\eta > t$, the $\Omega_t$Integral can be calculated using the 
machineries of Stoke's theorem, Cauchy integral, and the Schwarz
function of the ellipse.
\begin{eqnarray*}
 & \dint_{\Omega_t}\dfrac{d^2\xi}{z - \xi}
  = \dfrac{1}{2i} \dint_{\Omega_t}\dfrac{d\bar \xi d\xi}{z - \xi}
 = \dfrac{1}{2i} \dint_{\Omega_t} d \left(\dfrac{\bar \xi d\xi}{z - \xi}\right)
 = \dfrac{1}{2i} \dint_{\partial\Omega_t}\dfrac{\bar \xi d\xi}{z - \xi}
 \\
 & = \dfrac{1}{2i} \dint_{\partial\Omega_t}
      \dfrac{S_1(\xi) + S_2(\xi)}{z - \xi}\, d\xi
      = \pi S_2(z) 
 = \dfrac{2\pi ab}{c^2}(z - (z^2-c^2t^2)^{1/2}) 
\end{eqnarray*}
Thus the ESL outside lens equation is
\opeqn
 \omega = z - \dint_0^1 \dfrac{\tilde\sigma(t) t dt}{(\bar z^2-c^2t^2)^{1/2}} \ ,
\cleqn
and the ESL inside lens equation is
\opeqn
 \omega = z - \dint_0^\eta \dfrac{\tilde\sigma(t) t dt}{(\bar z^2-c^2t^2)^{1/2}} 
\cleqn
where $c$ is the focal length of the mass boundary ellipse
and $\eta$ is elliptic radial parameter of $z$.

\subsection{Inverse Elliptic Radial Density: $\tilde\sigma(t) = 1/t$}

If the mass density is proportional to the inverse of the elliptic
radius, $\tilde\sigma(t) =1/t$, then the outside and inside 
lens equations are
\opeqn
  \omega = z - \frac{1}{c}\arcsin \left(\dfrac{c}{\bar z}\right) 
     \ : \qquad
     z\ {\rm is \ outside\ the\ lens\ mass} 
\label{eqLeqOutside}
\cleqn  
and 
\opeqn
 \omega = z - \dfrac{1}{c} \arcsin \left(\dfrac{c\eta}{\bar z}\right)
   \ : \qquad
   z \ {\rm is\ inside\ the\ lens\ mass}, \ z \neq 0  
\label{eqLeqInside}
\cleqn
where $\eta$ is the elliptic radius of $z$: 
$z = \eta(a\cos\theta + i b \sin\theta)$.
We call the ESL lens with $\tilde\sigma(t) = 1/t$ an arcsin lens
named after the arcsin functions of the deflection angles.

\subsection{Constant Density: $\tilde\sigma(t)=2$}

If the projected mass density is constant, $\tilde\sigma(t)=2$.
The outside equation is
\opeqn
  \omega = z-  \dfrac{2}{c^2}(\bar z - (\bar z^2 - c^2)^{1/2} ) \ ,
\cleqn
and the inside equation is 
\opeqn
\omega = z-  \dfrac{2}{c^2}( \bar z - (\bar z^2 - c^2\eta^2)^{1/2} ) 
              = z - \dfrac{1}{ab} \left(z - \dfrac{a-b}{a+b}\bar z\right)
 \cleqn
 where the last equality holds because $\bar z = S(z)$ on 
 the $\eta$-ellipse.

\section{Points and Sticks and Disks}

A set of lens masses that are either 
spherically symmetric or ellliptically symmetric can be 
represented by a set of points and sticks for the images outside
the lensing masses. If we normalize the lens equation by the total
mass, then the normalized outside lens equation is given by
\opeqn
 \omega = z - \Sigma_j \dfrac{\epsilon_{cj}}{\bar z - \bar x_{cj}} 
            - \Sigma_k \dint_0^1\dfrac{\epsilon_{ek}\, 
                       \tilde\sigma(t)\, t dt}
      {((\bar z - \bar x_{ek})^2 - e^{i2\theta_k} c_k^2 t^2)^{1/2}}
     \ ;
\label{eqLeqMaster}
\cleqn  
\opeqn
    \Sigma_j \epsilon_{cj} + \Sigma_k \epsilon_{ek} = 1 \ .
\cleqn
$\epsilon_{cj}$ and $x_{cj}$ are the fractional mass and position
of the center of mass of the $j$-th point (circularly symmetric mass); 
$\epsilon_{ek}$, $x_{ek}$, 
$c_k$ and $\theta_k$ are the fractional mass, position of the center
of mass, focal length and direction angle of  
the $k$-th stick (elliptically symmetric mass).  
The focal points of the $k$-th elliptic mass are at
$z = x_{ek} \pm e^{i\theta_k} c_k$.

For an image inside the disk of a circularly symmetric lens $j_0$, the 
lens equation is obtained by replacing the $j_0$-th point mass lens
deflection angle in eq.(\ref{eqLeqMaster}) by its inside deflection angle.
\opeqn
 \dfrac{\epsilon_{cj_0}}{\bar z - \bar x_{cj_0}} \quad \Rightarrow \quad
 \dfrac{\epsilon_{cj}\eta[z-x_{cj_0}]}{\bar z - \bar x_{cj}} 
\cleqn
where $\eta[z-x_{cj_0}]$ denotes the elliptic radial parameter of 
$z-x_{cj_0}$.

For an image inside the disk of an elliptically symmetric lens $k_0$,
the lens equation is obtained by replacing the $k_0$-th elliptical
mass lens deflection angle in 
eq.(\ref{eqLeqMaster}) by its inside deflection angle by changing
the upper bound of the integral from $1$ to $\eta[z-c_{ek_0}]$.

\vskip 0.8cm

We thank Dmitri Khavinson for discussions, and Alexandre Eremenko for 
a critical comment that helps to delineate our finding.
We appreciate the Rhie family for their financial support.

\clearpage



\end{document}